\documentclass[aps,pra,twocolumn,showpacs]{revtex4}

\usepackage[dvips]{graphicx}
\usepackage[normalem]{ulem}
\usepackage{amsmath}

\begin{document}

\title{Density profiles of polarized Fermi gases confined in harmonic traps}

\author{G. Bertaina, and S. Giorgini}
\affiliation{Dipartimento di Fisica, Universit\`a di Trento and CRS-BEC INFM, I-38050 Povo, Italy}

\date{\today}

\begin{abstract}
On the basis of the phase diagram of the uniform system we calculate the density profiles of a trapped polarized Fermi gas at zero temperature using the local density approximation. By varying the overall polarization and the interaction strength we analyze the appearance of a discontinuity in the profile, signalling a first order phase transition from a superfluid inner core to a normal outer shell. The local population imbalance between the two components and the size of the various regions of the cloud corresponding to different phases are also discussed. The calculated profiles are quantitatively compared with the ones recently measured by Shin {\it et al.}, Phys. Rev. Lett. {\bf 101}, 070404 (2008).       
\end{abstract}


\maketitle

\section{Introduction}
\label{Introduction}

The field of ultracold two-component Fermi gases with population imbalance is a very active area of research which in recent years has attracted a great deal of interest both experimentally and theoretically~\cite{RMP08}. The experiments are carried out with clouds of atoms confined in harmonic traps and the analysis of the measured density profiles is a key diagnostic tool to investigate issues such as shell structures and the phase transition from a superfluid to a normal state of the gas. Important achievements of this technique have been the observation of phase separation between a superfluid core and a normal external shell driven by the degree of polarization~\cite{Zwierlein06a,Partridge06a}, by the interaction strength tuned across a Feshbach resonance~\cite{Shin06,Shin08} and by the temperature of the gas~\cite{Zwierlein06b,Partridge06b,Shin08nat}. 

In the unitary limit, corresponding to a diverging scattering length between the spin-up and spin-down components of the fermionic mixture, the density profiles of the polarized gas have been investigated in details in Ref.~\cite{Lobo06}. This theoretical study is based on an accurate determination of the equation of state of the strongly-interacting normal gas as a function of imbalance obtained using quantum Monte Carlo methods. It provides predictions for the shape of the profiles, for the density jump at the boundary of the superfluid core and for the critical polarization when the system turns fully normal, which are in excellent agreement with the experimental findings of the MIT group~\cite{Shin08nat}. In the Bose-Einstein condensate (BEC) limit of small and positive scattering lengths, the polarized gas is predicted to behave like a mixture of composite bosons (the bound dimer molecules) and fermions (the unpaired atoms)~\cite{Pieri06,Taylor07,Iskin08}. The density profiles in this regime have been investigated theoretically in Refs.~\cite{Pieri06,Iskin08} and experimentally in the recent study by the MIT group~\cite{Shin08}, where the Bose-Fermi mixture model is quantitatively tested at the level of mean-field theory, including also higher order corrections. At finite temperature, the density profiles of a trapped polarized Fermi gas have also been the object of theoretical investigations based on self-consistent approaches both at unitarity~\cite{Chien07} and in the BEC regime~\cite{Iskin08}.              

The phase diagram of the uniform gas at zero temperature as a function of polarization and interaction strength has been calculated using quantum Monte Carlo (QMC) techniques in Ref.~\cite{Pilati08}. This study provides a precise determination of the equation of state of four different phases competing for the ground state of the system: (a) the unpolarized and (b) polarized superfluid gas and (c) the fully and (d) partially polarized normal gas. The quantum phase transition from the normal to the superfluid state is first order and is accompanied by a region of phase separation where the two phases coexist in equilibrium. Only in the deep BEC regime, where the Bose-Fermi mixture model applies, the transition from a fully polarized Fermi gas to a miscible mixture of few superfluid bosons in a Fermi sea becomes second order. 

In this article we use the knowledge of the energy functionals of the uniform phases (a)-(d) as a function of polarization and interaction strength and using the local density approximation we calculate the phase diagram of a trapped gas and the density profiles of both spin components. We determine the conditions for the appearance of a superfluid unpolarized core and of a jump in the density profile signalling the occurrence of the first order quantum phase transition from superfluid to normal. The calculated density profiles are compared with the experimental ones of Ref.~\cite{Shin08} for different values of the interaction strength, from the unitary to the BEC limit, and for different values of the polarization. In Sec.~\ref{Section1} we introduce the equations of state of the four uniform phases (a)-(d) and in Sec.~\ref{Section2} we discuss the equations of mechanical and chemical equilibrium when the harmonic external potential is present. In Sec.~\ref{Section3} we present the results for the phase diagram, the density profiles and for the radii of the various shells present in the cloud. Conclusions are drawn in Sec.~\ref{Conclusions}.

\begin{figure}
\begin{center}
\includegraphics*[width=8.5cm]{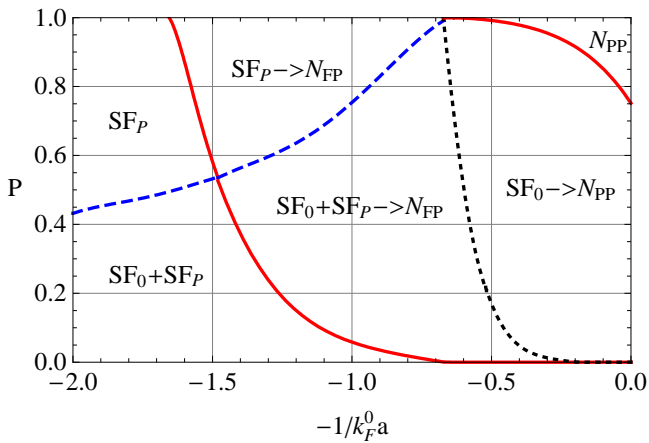}
\caption{(color online). Phase diagram for trapped atoms at zero temperature as a function of polarization and interaction strength. The regions correspond to different shell structures (see text). Inside the two (red) solid lines a jump in the density profile marks the superfluid-normal first order phase transition. Above the (blue) dashed line the unpolarized SF$_0$ superfluid inner core is absent. On the right of the dotted line the polarized SF$_{\text P}$ superfluid phase is absent, while on the left the partially polarized N$_{\text{PP}}$ normal phase is absent.}
\label{fig1}
\end{center}
\end{figure}

\begin{figure*}
\scalebox{.95}
{\begin{tabular}{ccccc}
\includegraphics[angle=0, width=0.20\textwidth]{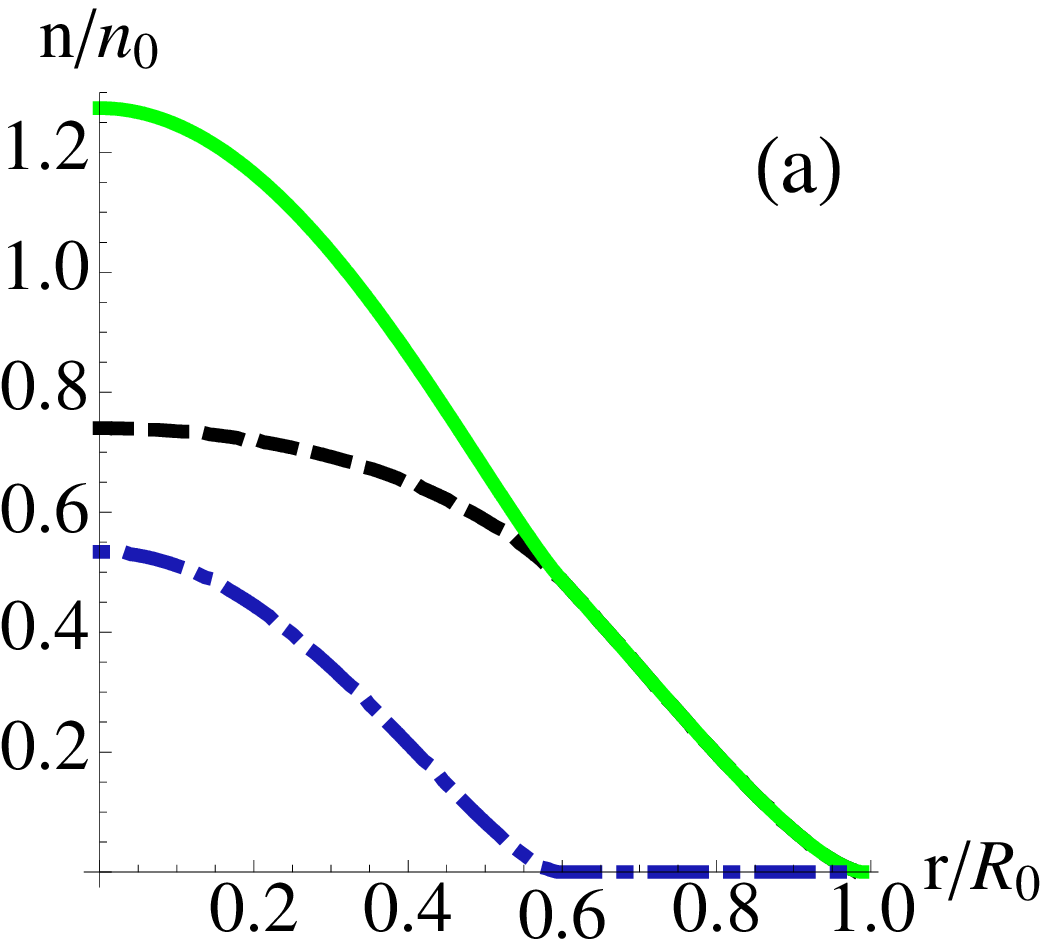}  &
\includegraphics[angle=0, width=0.20\textwidth]{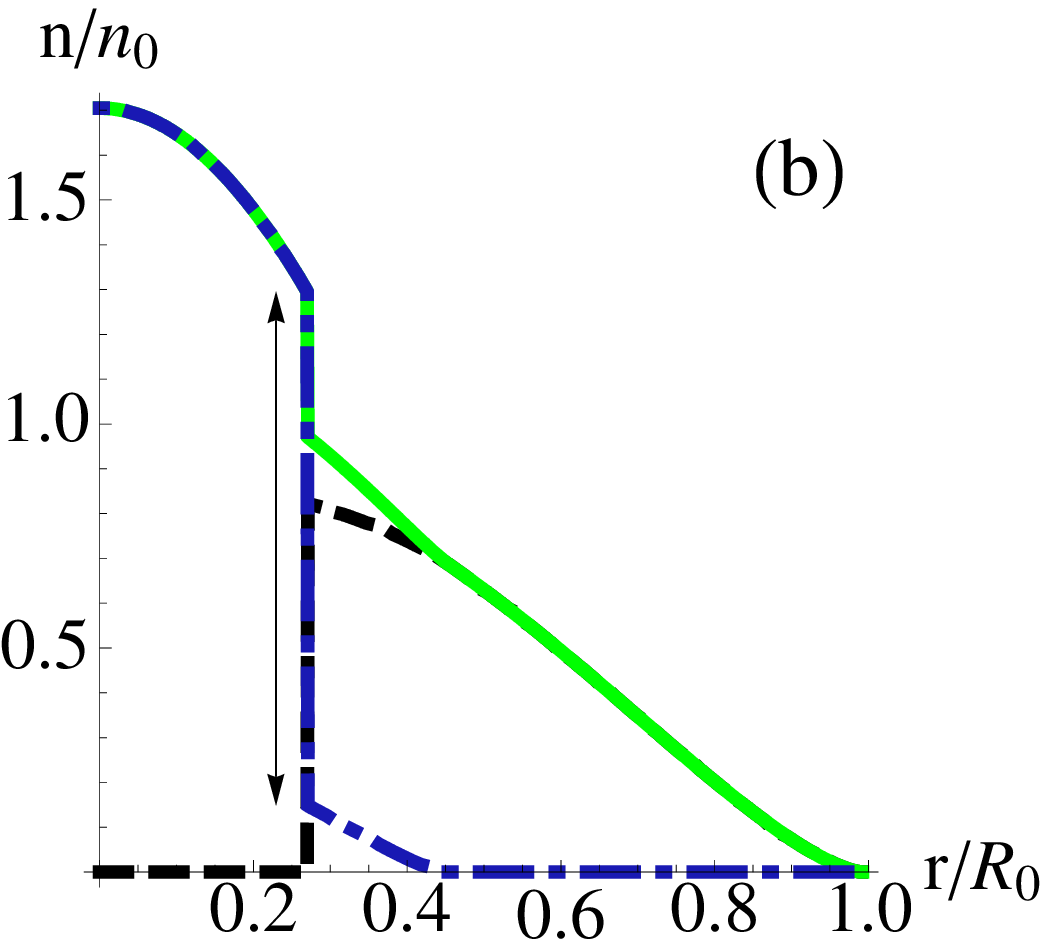}  &
\includegraphics[angle=0, width=0.20\textwidth]{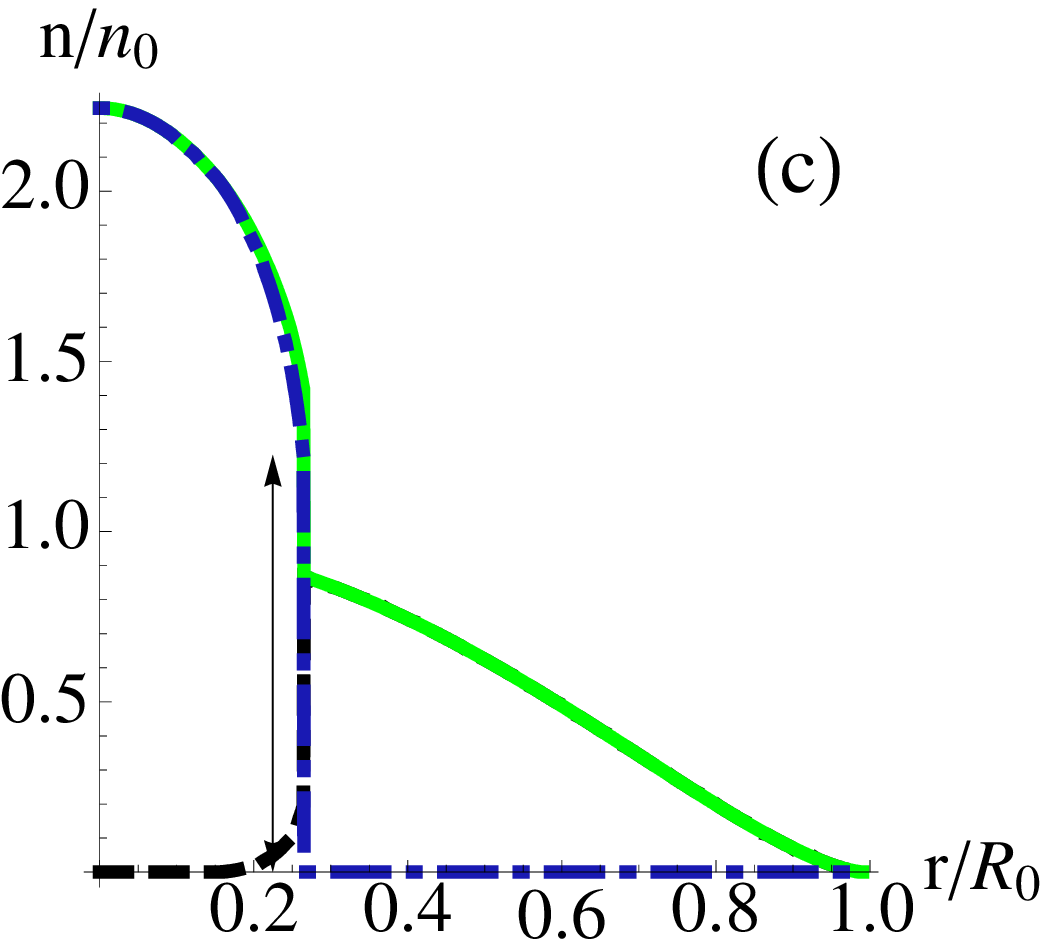}  &
\includegraphics[angle=0, width=0.20\textwidth]{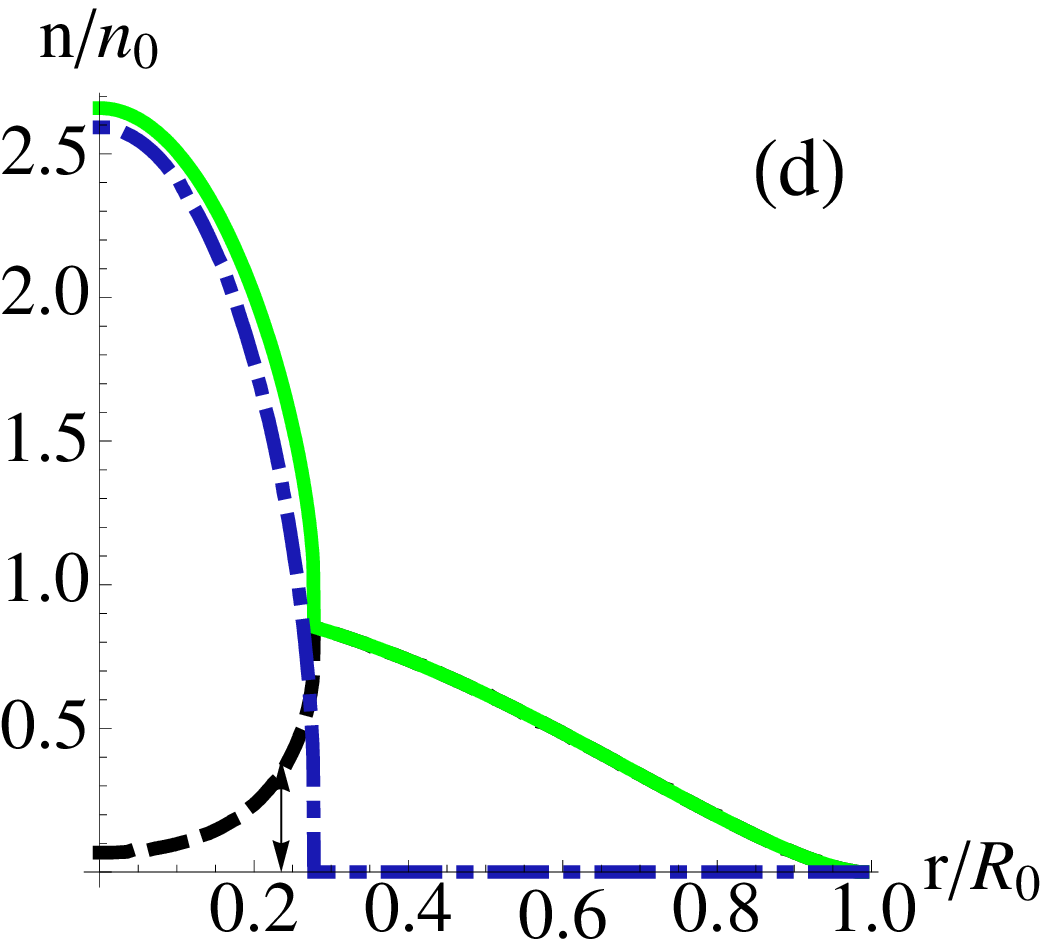}  &
\includegraphics[angle=0, width=0.20\textwidth]{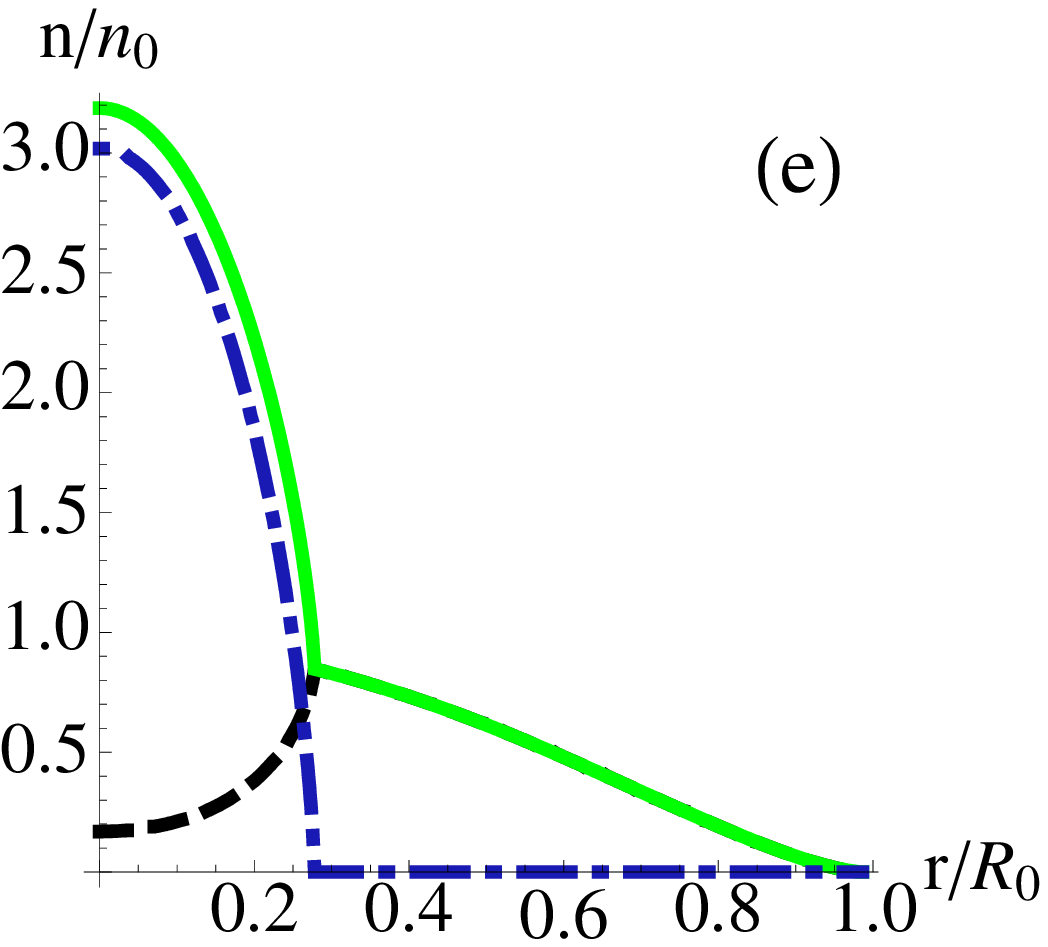}  \\
\includegraphics[angle=0, width=0.20\textwidth]{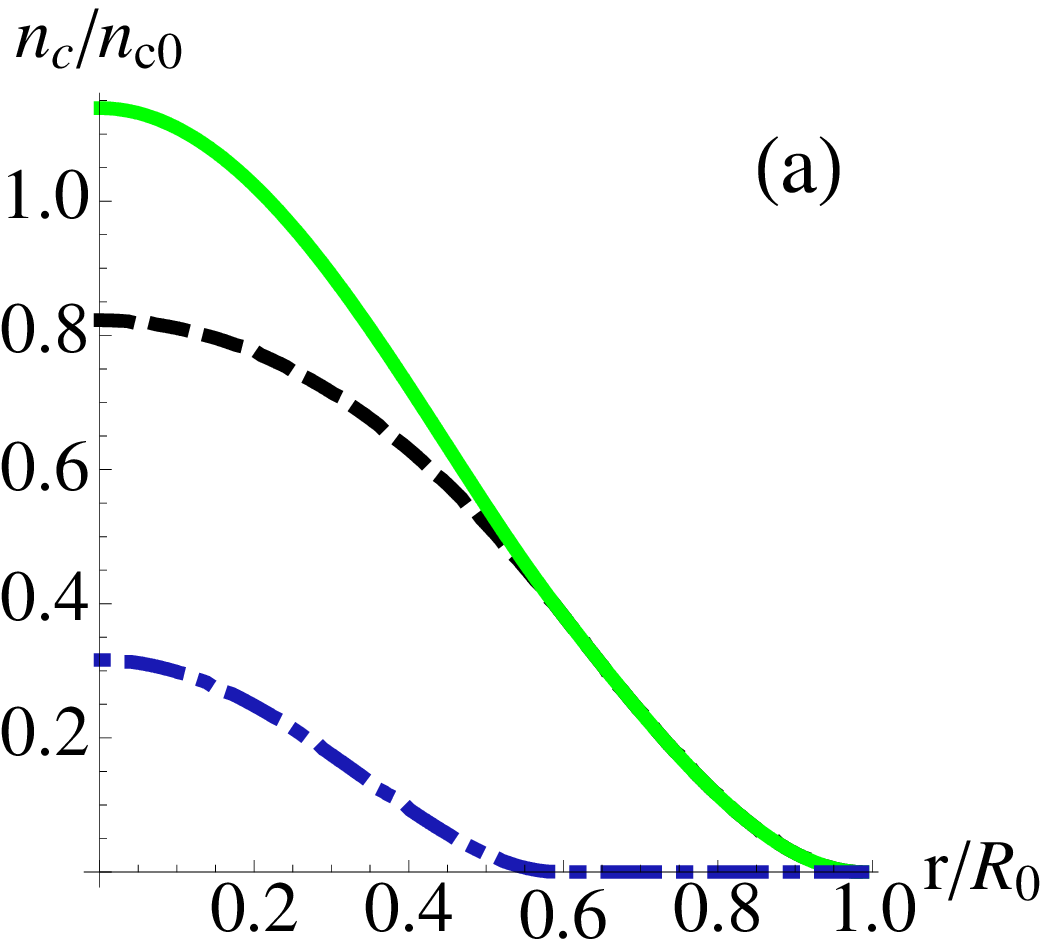} &
\includegraphics[angle=0, width=0.20\textwidth]{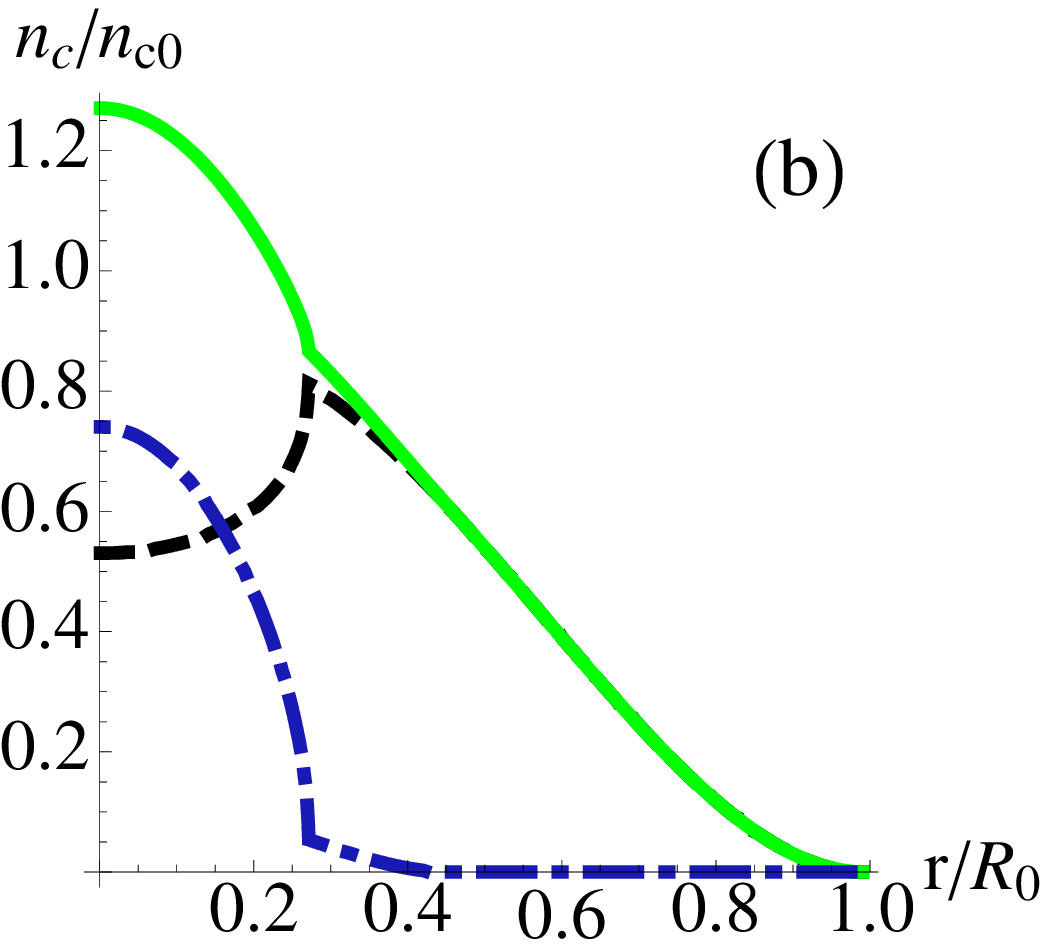} &
\includegraphics[angle=0, width=0.20\textwidth]{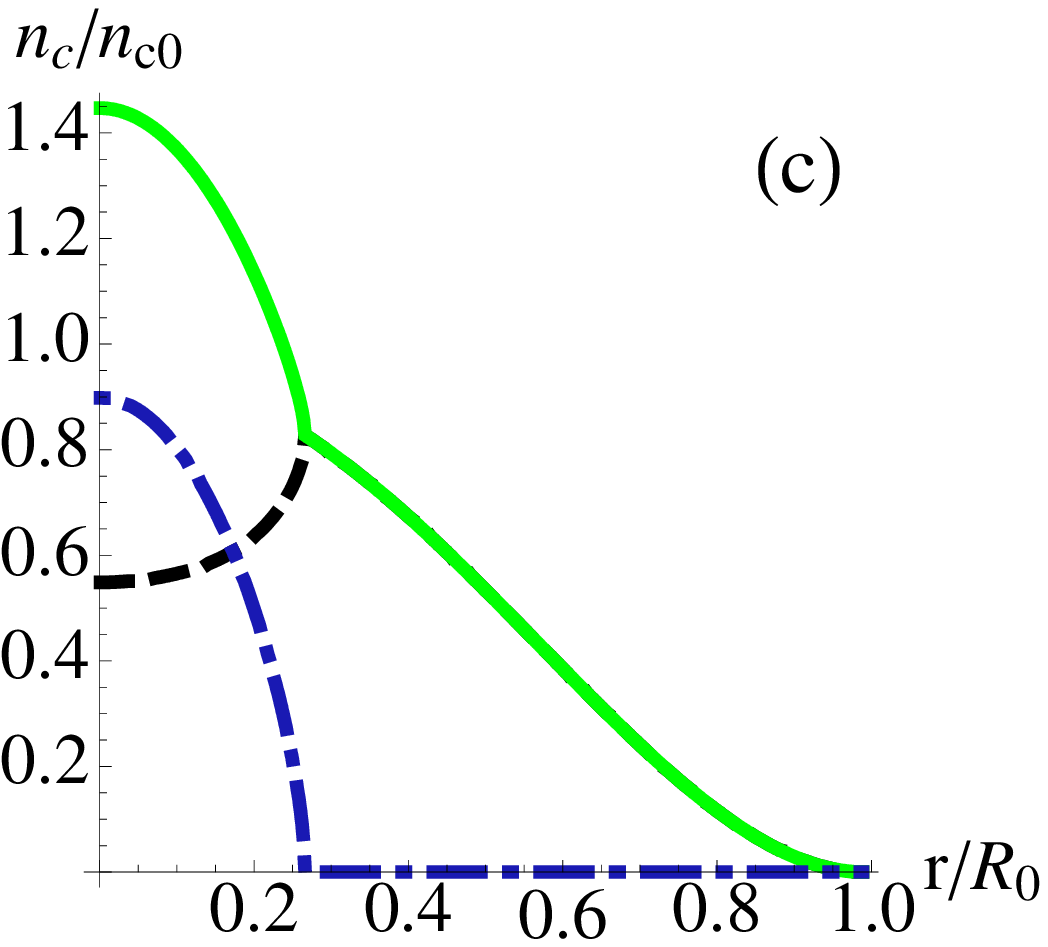} &
\includegraphics[angle=0, width=0.20\textwidth]{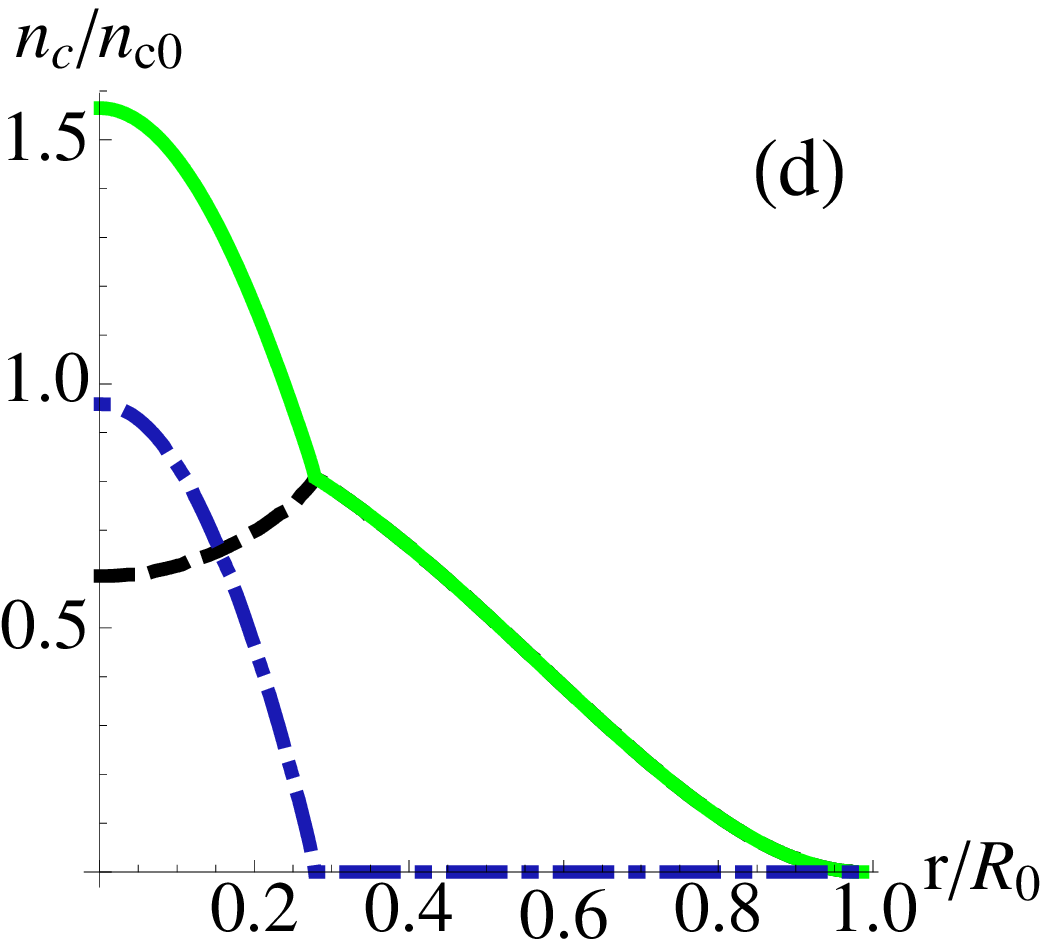} &
\includegraphics[angle=0, width=0.20\textwidth]{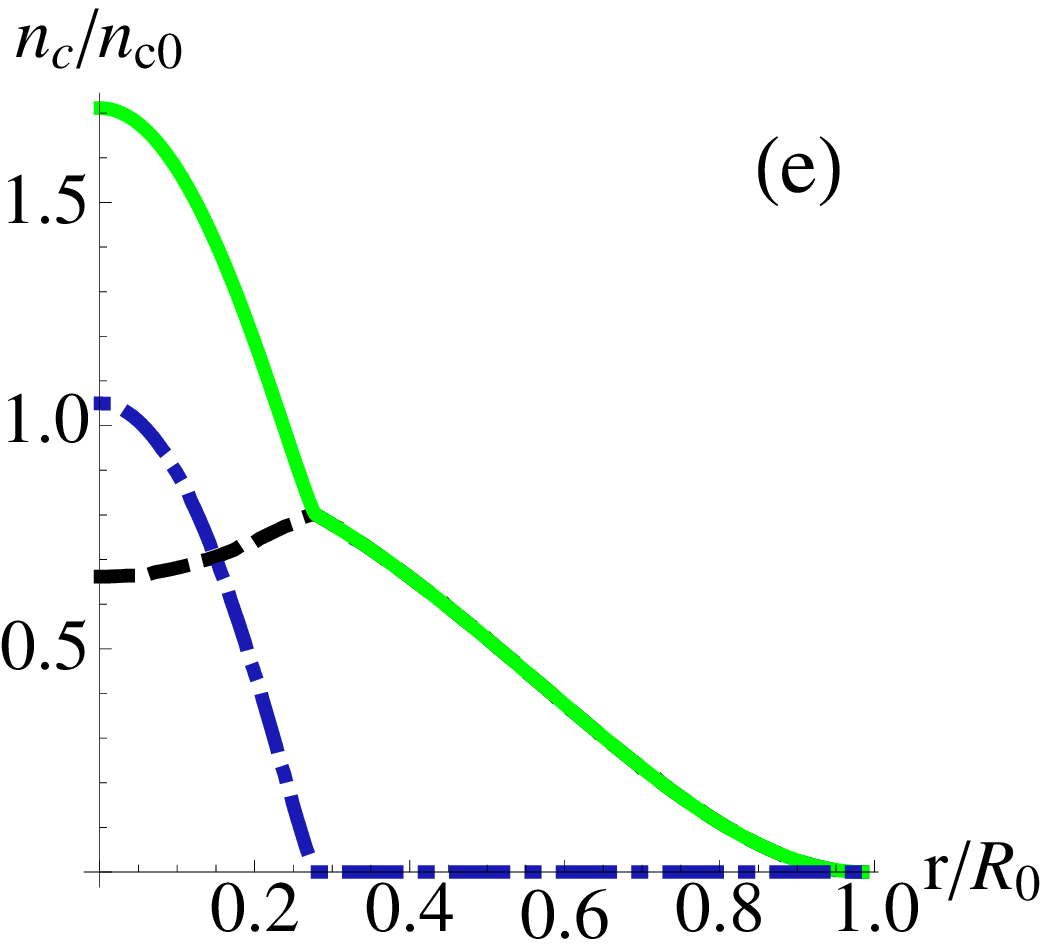} \\
\end{tabular}}
\caption{(color online). Density (upper row) and column density (lower row) profiles at $P=0.8$ for different values of the interaction strength. Column (a) corresponds to unitarity ($1/k_{F\uparrow}^0a=0$), column (b) to $1/k_{F\uparrow}^0a=0.4$, column (c) to $1/k_{F\uparrow}^0a=0.8$, column (d) to $1/k_{F\uparrow}^0a=1.2$ and column (e) to $1/k_{F\uparrow}^0a=1.7$. The solid (green) lines refer to the majority spin-up component, the dot-dashed (blue) lines to the minority spin-down component and the dashed (black) lines to the density difference $n_\uparrow-n_\downarrow$. The jump in the density $n_\uparrow$ of the minority component is shown with a vertical double arrow. The reference density $n_0$, column density $n_{c0}$ and radius $R_0$ all refer to a non-interacting Fermi gas with $N_\uparrow$ particles.}
\label{fig2}
\end{figure*}

\section{Uniform phases}
\label{Section1}

In this Section we introduce the uniform phases considered in Ref.~\cite{Pilati08} and we discuss their energy densities as a function of the concentration of spin-down particles (the minority species) and of the interaction strength. The natural unit of energy is provided by the Fermi energy of the spin-up majority component
\begin{equation}
E_{F\uparrow}=\frac{\hbar^2k_{F\uparrow}^2}{2m}=\frac{\hbar^2(6\pi n_\uparrow)^{2/3}}{2m} \;,
\label{EF}
\end{equation}
where $k_{F\uparrow}$ is the Fermi wave vector of the spin-up particles fixed by their particle density $n_\uparrow$. For trapped systems, where the density varies with position, the above equation defines the local Fermi energy. 

(a) {\it Unpolarized superfluid gas} (SF$_0$). In this phase the spin-up and spin-down particle densities are equal $n_\uparrow=n_\downarrow=n/2$. For positive values of the $s$-wave scattering length ($a>0$) the energy density can be written in the form
\begin{equation}
{\cal E}_{\text{SF}_0}(n/2)=\frac{n}{2}\epsilon_b + \left(\frac{3}{5}n_\uparrow E_{F\uparrow}\right) 
2G(1/k_{F\uparrow}a) \;.
\label{EOS1}
\end{equation} 
In the above equation $\epsilon_b$ denotes the binding energy of the molecule ($\epsilon_b=-\hbar^2/ma^2$ for a zero-range potential) and $G(1/k_{F\uparrow}a)$ is a dimensionless function of the interaction parameter $\eta=1/k_{F\uparrow}a$. A suitable parametrization of $G(\eta)$, which well reproduces the QMC results from the BEC to the unitary limit (see Ref.~\cite{Pilati08}), is provided by the following formula
\begin{equation}
G(\eta)=\left\{  \begin{array}{cc} \alpha_0+\alpha_1 \arctan(\alpha_2\eta)   & (\eta<0.699)\;,  \\
                                   \epsilon(\eta)+\frac{\alpha_3}{\eta^3} +\frac{\alpha_4}{\eta^4} & (\eta>0.699)\;,
                                   \end{array} \right.
\label{EOS11}
\end{equation}
where we defined the function $\epsilon(\eta)=5(0.60/\eta)/(18\pi)[1+128(0.60/\eta)^{3/2}/(15\sqrt{6}\pi^{3/2})]$, while the constants $\alpha_{0-4}$ have the following values: $\alpha_0=0.434$, $\alpha_1=-0.292$, $\alpha_2=2.90$, $\alpha_3=0.0129$ and $\alpha_4=-0.0100$. The function $G$ as defined above is continuous at $\eta=0.699$ with continuous first derivative. For $k_{F\uparrow}a\ll 1$ it reduces to the function $\epsilon(\eta)$, which corresponds to the energy, including the Lee-Huang-Yang (LHY) beyond mean-field correction, of a gas of composite bosons interacting with the dimer-dimer scattering length $a_{dd}=0.60a$.

(b) {\it Polarized superfluid gas} (SF$_{\text P}$). This phase is characterized by a density $n_P=n_\downarrow$ of pairs and a density $n_A=n_\uparrow-n_\downarrow$ of unpaired atoms, such that $n=2n_P+n_A$. The concentration of the minority atoms is denoted by $y=n_\downarrow/n_\uparrow$. For $1/k_{F\uparrow}a\geq0.5$ an accurate parametrization of the equation of state is provided by the following energy functional  
\begin{eqnarray}
{\cal E}_{\text{SF}_{\text {P}}}&=&{\cal E}_{\text{SF}_0}(n_P) 
\label{EOS2} \\
&+&\frac{3}{5}n_\uparrow E_{F\uparrow} \left[(1-y)^{5/3} + \frac{5k_{F\uparrow}a_{ad}}{3\pi}y(1-y) \right]\;.
\nonumber
\end{eqnarray}
The first term on the right hand side is the energy density (\ref{EOS1}) of an unpolarized superfluid with density $2n_P$. The other two terms correspond to the kinetic energy of a gas with density $n_A$ of unpaired fermions and to the interaction energy, treated at the level of mean field, between unpaired and paired atoms parametrized by the atom-dimer scattering length $a_{ad}=1.18a$. With the above energy functional one recovers the Bose-Fermi mixture model in the deep BEC limit corresponding to $1/k_{F\uparrow}a\gg 1$.   

(c) {\it Fully polarized normal gas} (N$_{\text{FP}}$). In this phase the density of spin-down particles vanishes ($n_\downarrow=0$) and the energy density coincides with the one of an ideal Fermi gas
\begin{equation}
{\cal E}_{\text{N}_{\text{FP}}}=\frac{3}{5}n_\uparrow E_{F\uparrow} \;.
\label{EOS3}
\end{equation}
 
(d) {\it Partially polarized normal gas} (N$_{\text{PP}}$). This phase is characterized by the concentration $x=n_\downarrow/n_\uparrow$ of the spin-down particles. For small concentrations ($x\ll1$) the dependence on $x$ of the energy functional can be written in the form of the Landau-Pomeranchuk Hamiltonian of weakly interacting fermionic quasiparticles \cite{Baym91,Lobo06,Chevy06,Bulgac07,Combescot07,Pilati08,Prokofev08}
\begin{equation}
{\cal E}_{\text{N}_{\text{PP}}}=\frac{3}{5}n_\uparrow E_{F\uparrow} \left(1-Ax+\frac{m}{m^\ast}x^{5/3}+Fx^2\right)\;,
\label{EOS4}
\end{equation}
where $A$ is the binding energy of a single spin-down quasiparticle in the Fermi sea of spin-up particles, $m^\ast$ is its effective mass and $F$ represents the interaction between quasiparticles. These quantities all depend on $\eta=1/k_{F\uparrow}a$. We used the following parametrizations, which well reproduce the QMC results when $\eta>0$:
\begin{multline}
A(\eta)=-\frac{5\epsilon_b}{3E_{F\uparrow}}+\beta_0+\beta_1\eta+\beta_2\eta^2+\beta_3\eta^3 \;, 
\label{EOSA}
\end{multline}
where the first term on the right hand side is conveniently introduced in analogy with \eqref{EOS1} and the constants $\beta_{0-3}$ have the following values: $\beta_0=0.986$, $\beta_1=1.11$, $\beta_2=-2.23$, $\beta_3=0.847$;
\begin{equation}
\frac{m^\ast(\eta)}{m}=\gamma_0 + \gamma_1\eta + \gamma_2\eta^2 + \gamma_3\eta^3 + \gamma_4\eta^4  \;,
\label{EOSM}
\end{equation}
where the constants $\gamma_{0-4}$ are given by $\gamma_0=1.11$, $\gamma_1=0.255$, $\gamma_2=0.337$, $\gamma_3=0.0579$, $\gamma_4=-0.128$; and finally
\begin{equation}
F(\eta)= \delta_0 e^{-(\eta-\delta_1)^2/\delta_2}  \;,
\label{EOSF} 
\end{equation}
where the constants $\delta_{0-2}$  have the values: $\delta_0=0.279$, $\delta_1=0.438$, $\delta_2=0.277$. The equation of state \eqref{EOS4} only holds for values of the interaction parameter $\eta\leq0.73$, since for larger values the normal gas becomes fully polarized with $x=0$ (see Ref.~\cite{Pilati08}). We also notice that the values of the effective mass $m^\ast$ are smaller than the exact diagrammatic Monte Carlo results of Ref.~\cite{Prokofev08} (see comment in Ref.~\cite{Pilati08}). This inaccuracy is compensated in Eq.~\eqref{EOS4} by the coefficient $F$ of the $x^2$ term, which finally provides a very accurate fit to the QMC results at finite concentration. However, the reported value of $F$ in Eq.~\eqref{EOSF} underestimates the interaction effects between quasiparticles.

\section{Local Density Approximation}
\label{Section2}

In the presence of the harmonic confinement
\begin{equation}
V_{ho}({\bf r})=\frac{1}{2}m\omega_x^2x^2+\frac{1}{2}m\omega_y^2y^2+\frac{1}{2}m\omega_z^2z^2 \;,
\label{VHO}
\end{equation}
the particle density depends on position and the concentration $n_\downarrow/n_\uparrow$ varies locally. The most useful parameters for the phase diagram are expressed in this case in terms of the total number $N_\uparrow$ ($N_\downarrow$) of spin-up (spin-down) particles. These are the overall polarization
\begin{equation}
P=\frac{N_\uparrow-N_\downarrow}{N_\uparrow+N_\downarrow} \;,
\label{POL}
\end{equation}
and the interaction parameter
\begin{equation}
\frac{1}{k_{F\uparrow}^0a}=\frac{1}{(6\pi n^0)^{2/3}a}=\frac{a_{ho}/a}{(48N_\uparrow)^{1/6}} \;,
\end{equation}
where $a_{ho}=\sqrt{\hbar/m(\omega_x\omega_y\omega_z)^{1/3}}$ is the harmonic oscillator length and $k_{F\uparrow}^0$ and $n^0=n_\uparrow^0(0)$ are, respectively, the Fermi wave vector and the central trap density of a non-interacting Fermi gas with $N_\uparrow$ particles. 

In the local density approximation (LDA) the energy density of the inhomogeneous system is approximated at each spatial position by the one corresponding to a uniform gas at the local value of the density. This approximation becomes exact in the limit of large particle numbers, when the density changes slowly on the typical length scale associated with the external potential. We construct the grand-canonical potential $\Omega$ at zero temperature as the integral over space of the ground-state energy density $\cal{E}$, which depends on the local interaction parameter $\eta({\bf r})=[n_\uparrow({\bf r})/n^0]^{-1/3}/(k_{F\uparrow}^0a) $ and the local polarization $P({\bf r})=[n_\uparrow({\bf r})-n_\downarrow({\bf r})]/[n_\uparrow({\bf r})+n_\downarrow({\bf r})]$, that is
\begin{multline}
\Omega=\int d{\bf r} \left\{ {\cal E}\left[\eta({\bf r}),P({\bf r})\right] +[n_\uparrow({\bf r})
+n_\downarrow({\bf r})]V_{ho}({\bf r})\right.\\
\left. -n_\uparrow({\bf r})\mu_\uparrow-n_\downarrow({\bf r})\mu_\downarrow\right\} \;.
\label{grandcan}
\end{multline}
Here ${\cal E}$ corresponds to the uniform phase present at position ${\bf r}$, with the densities of the fermionic species replaced by the local densities, while $\mu_\uparrow$ and $\mu_\downarrow$ are the chemical potentials of the spin-up and spin-down particles, introduced as Lagrange multipliers. When studying the superfluid phase we introduce the chemical potential $\mu_S$ of the pairs and the chemical potential $\mu_A$ of the excess spin-up fermions; chemical equilibrium at the interface between the superfluid region and the normal region is imposed by setting $\mu_S=\mu_\uparrow+\mu_\downarrow$ and $\mu_A=\mu_\uparrow$. 

The density profiles are obtained by minimizing the grand-canonical potential \eqref{grandcan} with respect to the densities $n_\uparrow({\bf r})$ and $n_\downarrow({\bf r})$ of the two spin components. The condition of stationarity gives the LDA equations for the superfluid phase:
\begin{equation}
 \left\{\begin{array}{cc}
         \mu_S=&\partial{\cal E}_{\text{SF}_{\text {P}}}/\partial n_P+2V_{ho}({\bf r})\;,\\
         \mu_A=&\partial{\cal E}_{\text{SF}_{\text {P}}}/\partial n_A+V_{ho}({\bf r})\;,
         \end{array}\right.
\end{equation}
and for the normal phase:
\begin{equation}
 \left\{\begin{array}{cc}
        \mu_\uparrow=&\partial{\cal E}_{\text{N}_{\text {PP}}}/\partial n_\uparrow+V_{ho}({\bf r})\;,\\
        \mu_\downarrow=&\partial{\cal E}_{\text{N}_{\text {PP}}}/\partial n_\downarrow+V_{ho}({\bf r})\;.
        \end{array}\right.
\end{equation}
Notice that mechanical equilibrium is enforced by the continuity requirement of the grand-canonical potential density along the cloud profile. For each point in the harmonic trap we solve the two set of coupled equations, choosing the phase that minimizes the local grand-canonical potential and checking that the compressibility is positive. As usual, the integral of density over space allows us to fix the chemical potentials once we know the number of particles, or in our case, the total polarization and the interaction parameter.

\begin{figure}
{\begin{tabular}{cc}
\includegraphics[angle=0, width=0.47\columnwidth]{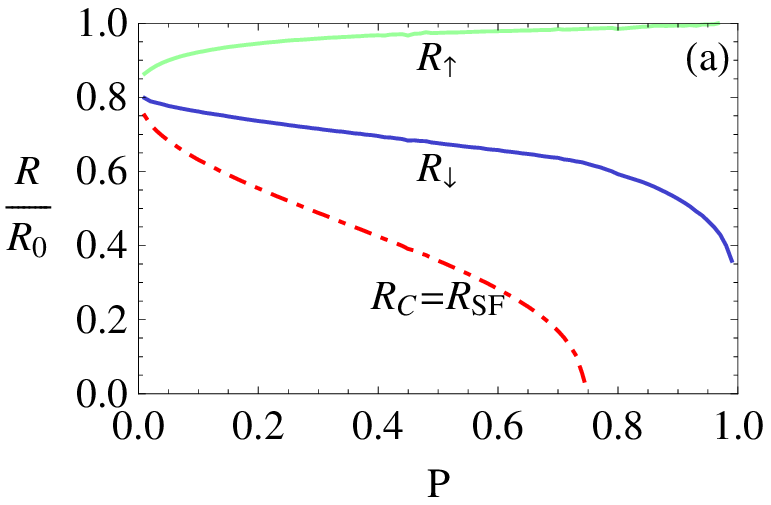}  &
\includegraphics[angle=0, width=0.47\columnwidth]{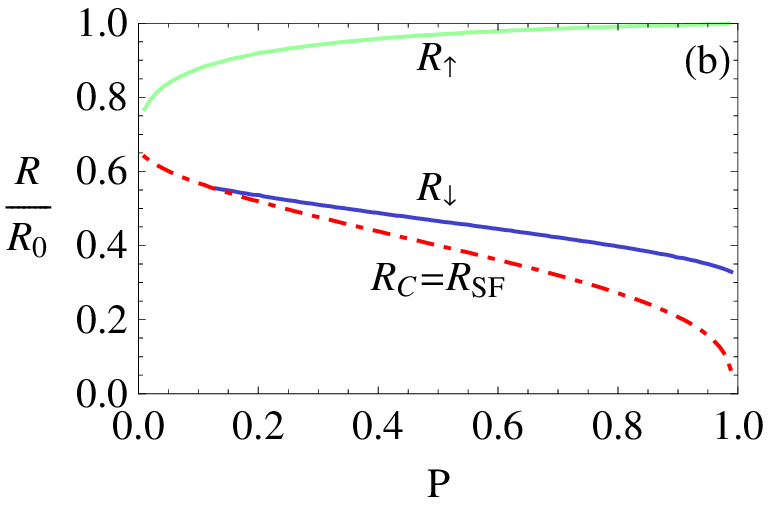} \\
\includegraphics[angle=0, width=0.47\columnwidth]{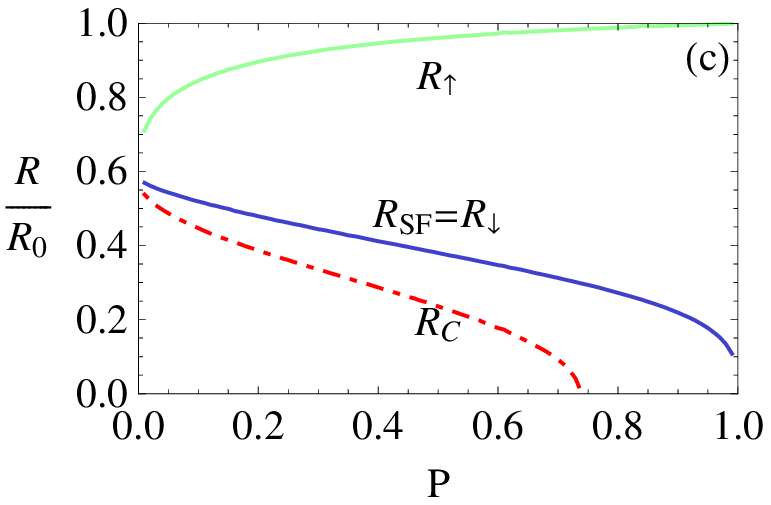}  &
\includegraphics[angle=0, width=0.47\columnwidth]{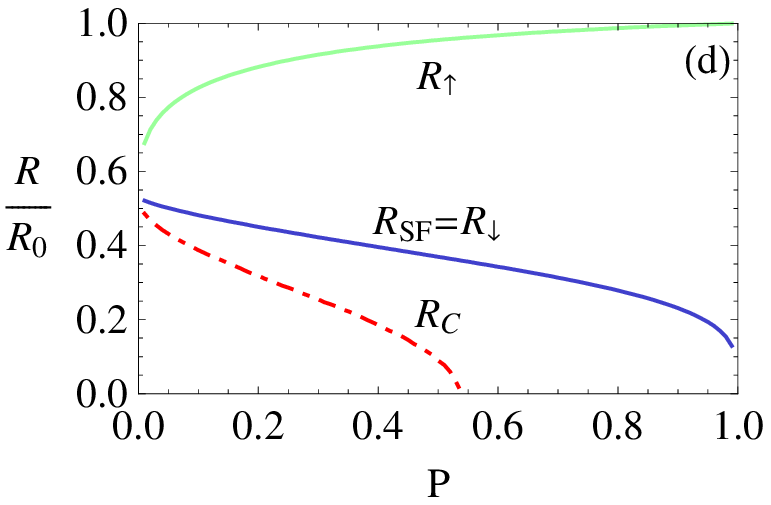} \\
\end{tabular}}
\caption{(color online). Radii of the spin-up ($R_\uparrow$) and spin-down ($R_\downarrow$) components and of the unpolarized superfluid core $R_c$ as well as of the superfluid shell $R_{\text{SF}}$ as a function of polarization for different values of the interaction strength. Panel: (a) $1/k_{F\uparrow}^0a=0$, (b) $1/k_{F\uparrow}^0a=0.5$, (c) $1/k_{F\uparrow}^0a=1$, (d) $1/k_{F\uparrow}^0a=1.5$. All the radii are divided by the radius $R_0$, which refers to a non-interacting Fermi gas with $N_\uparrow$ particles.}
\label{fig3}
\end{figure}

\begin{figure*}
\scalebox{.95}
{\begin{tabular}{ccccc}
\includegraphics[angle=0, width=0.20\textwidth]{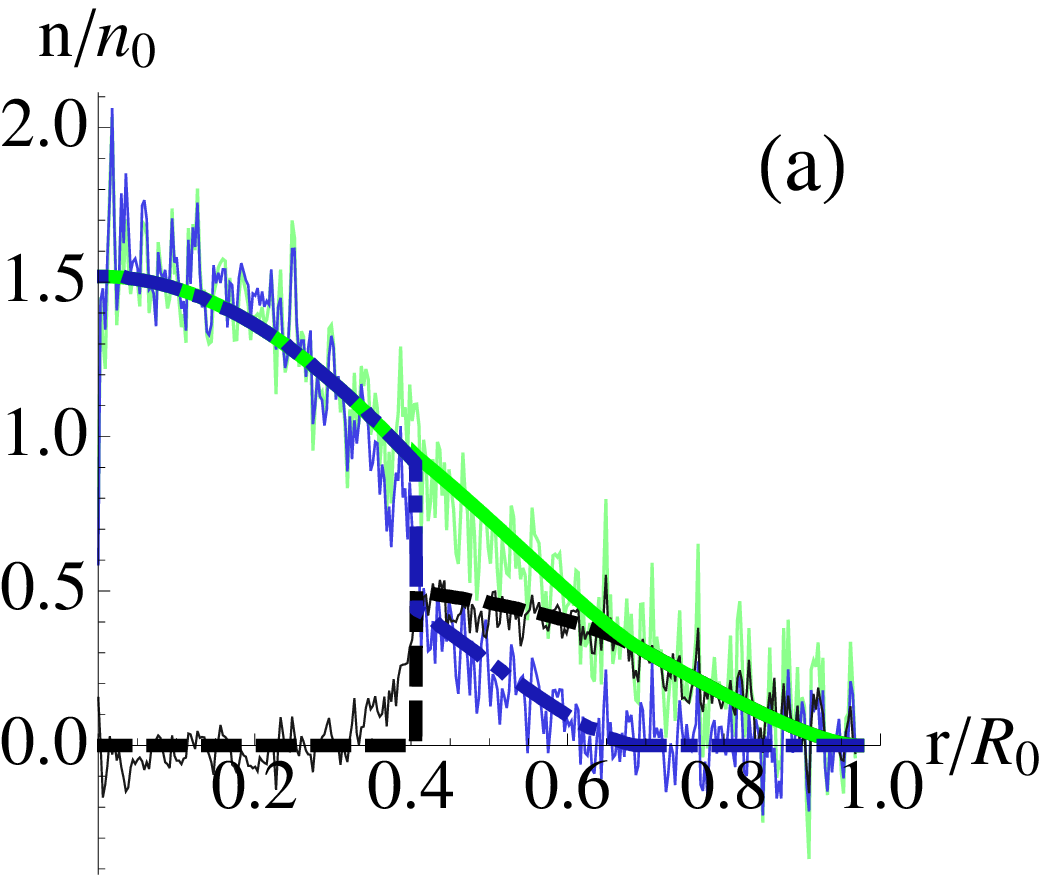}  &
\includegraphics[angle=0, width=0.20\textwidth]{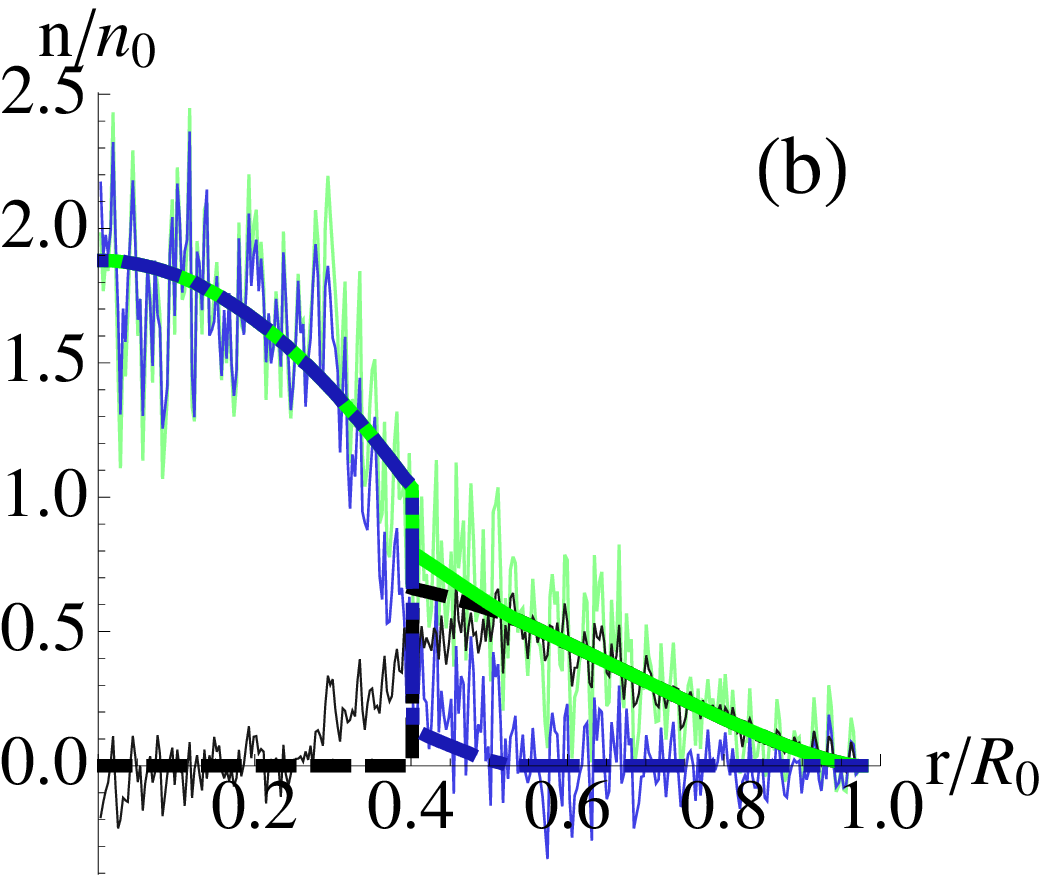}  &
\includegraphics[angle=0, width=0.20\textwidth]{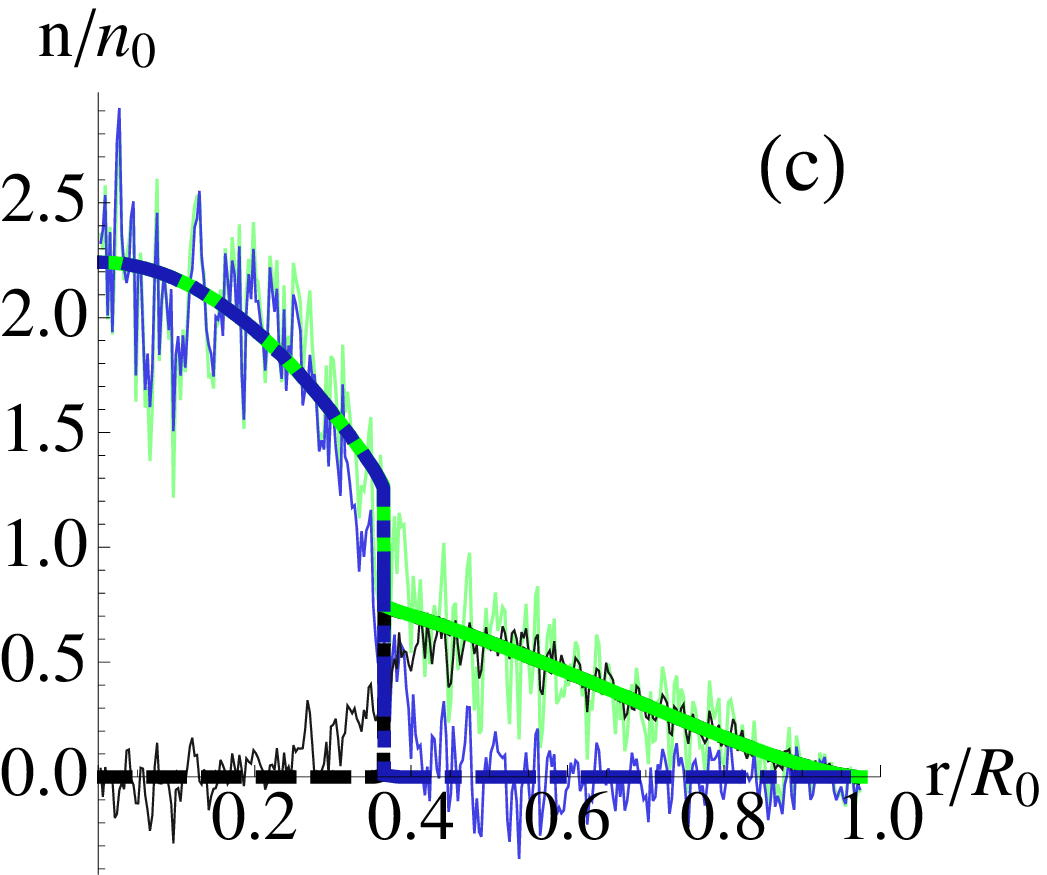}  &
\includegraphics[angle=0, width=0.20\textwidth]{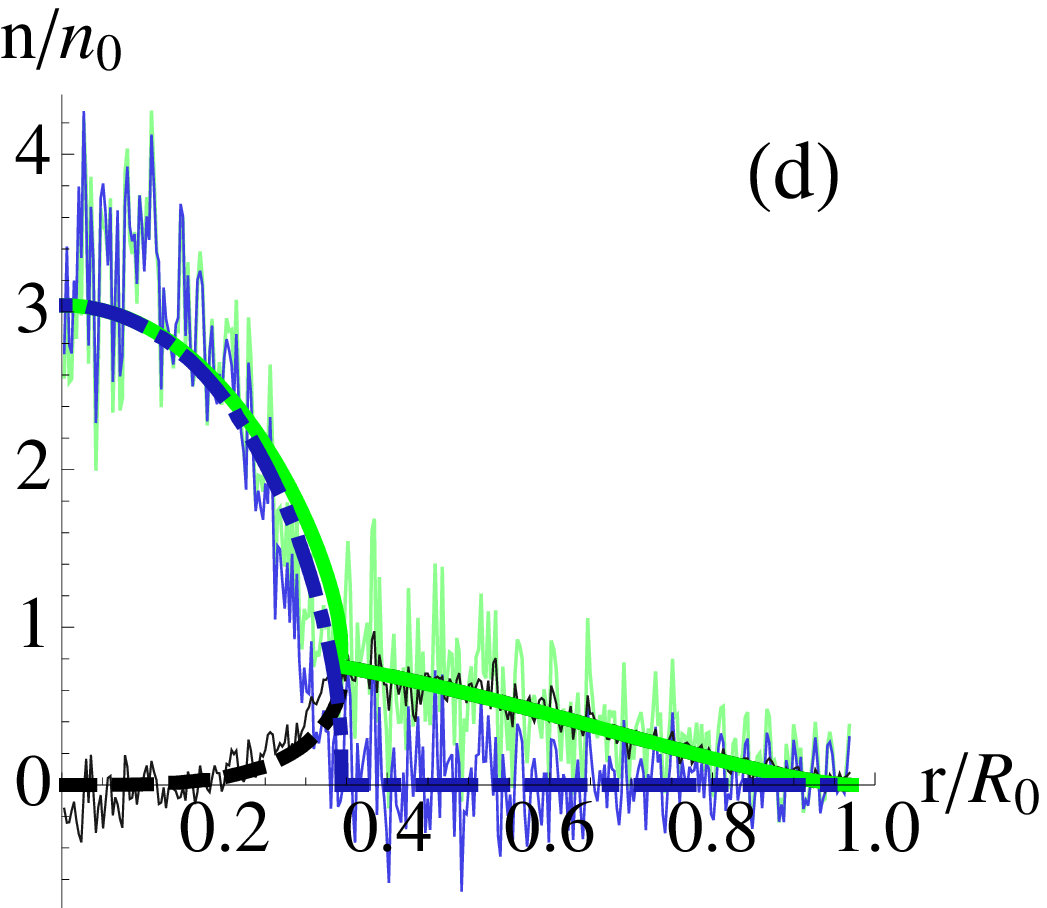}  &
\includegraphics[angle=0, width=0.20\textwidth]{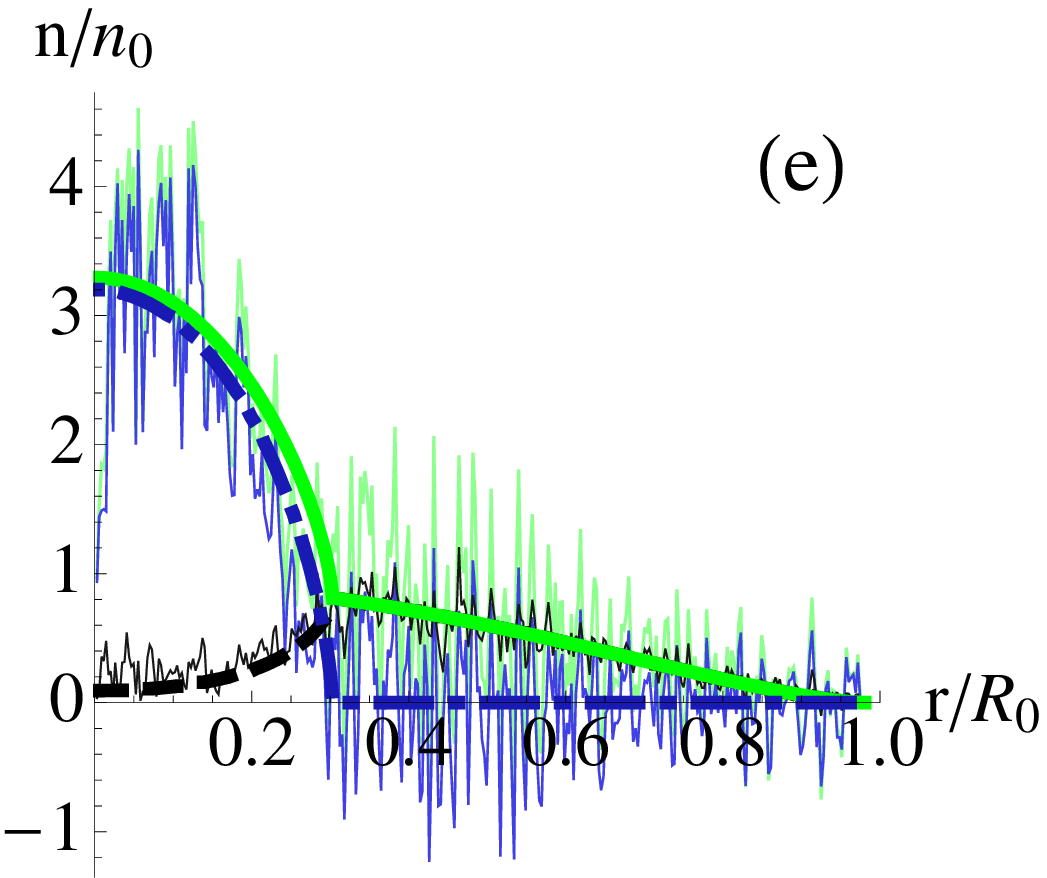}  \\
\includegraphics[angle=0, width=0.20\textwidth]{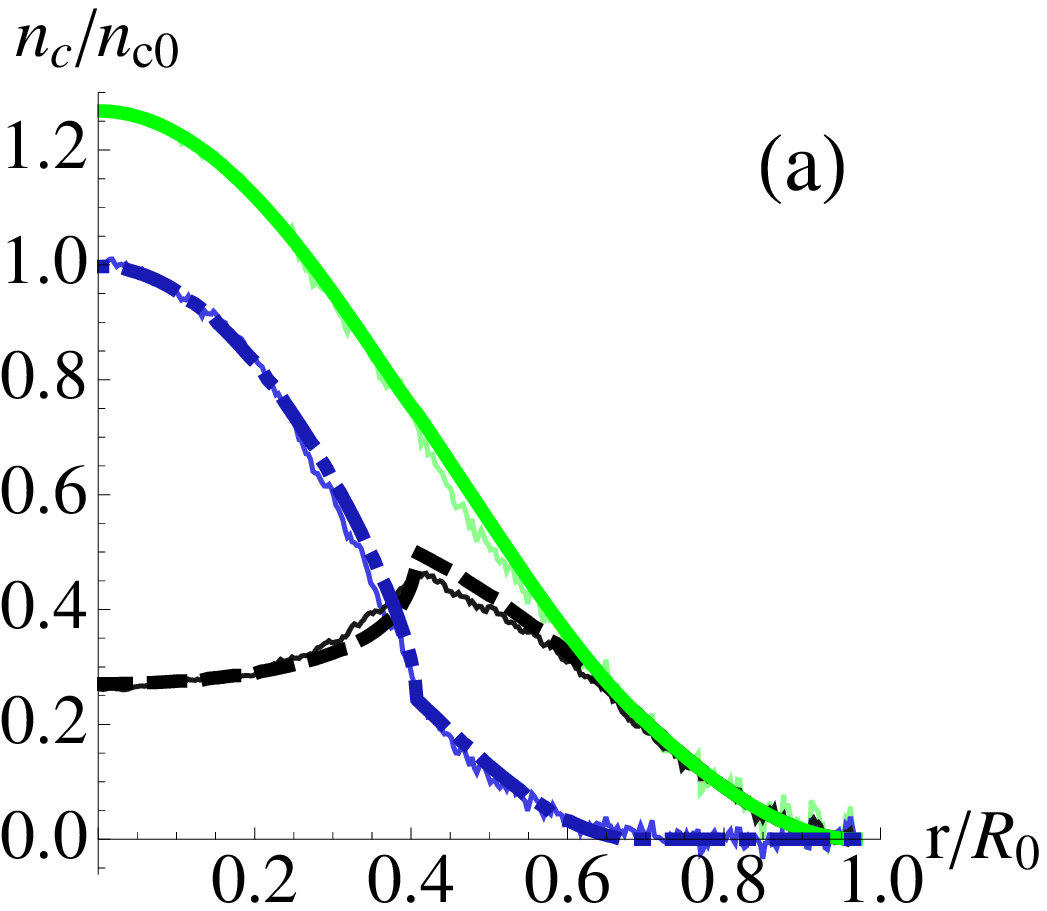} &
\includegraphics[angle=0, width=0.20\textwidth]{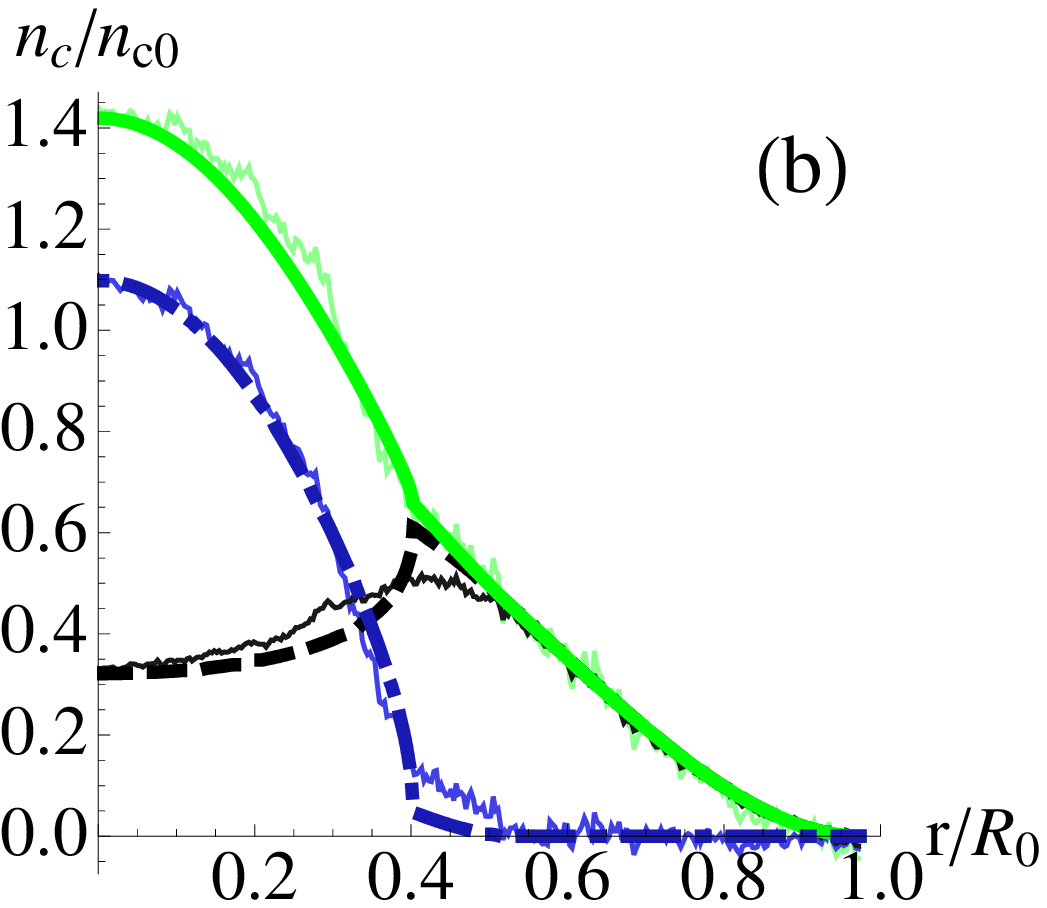} &
\includegraphics[angle=0, width=0.20\textwidth]{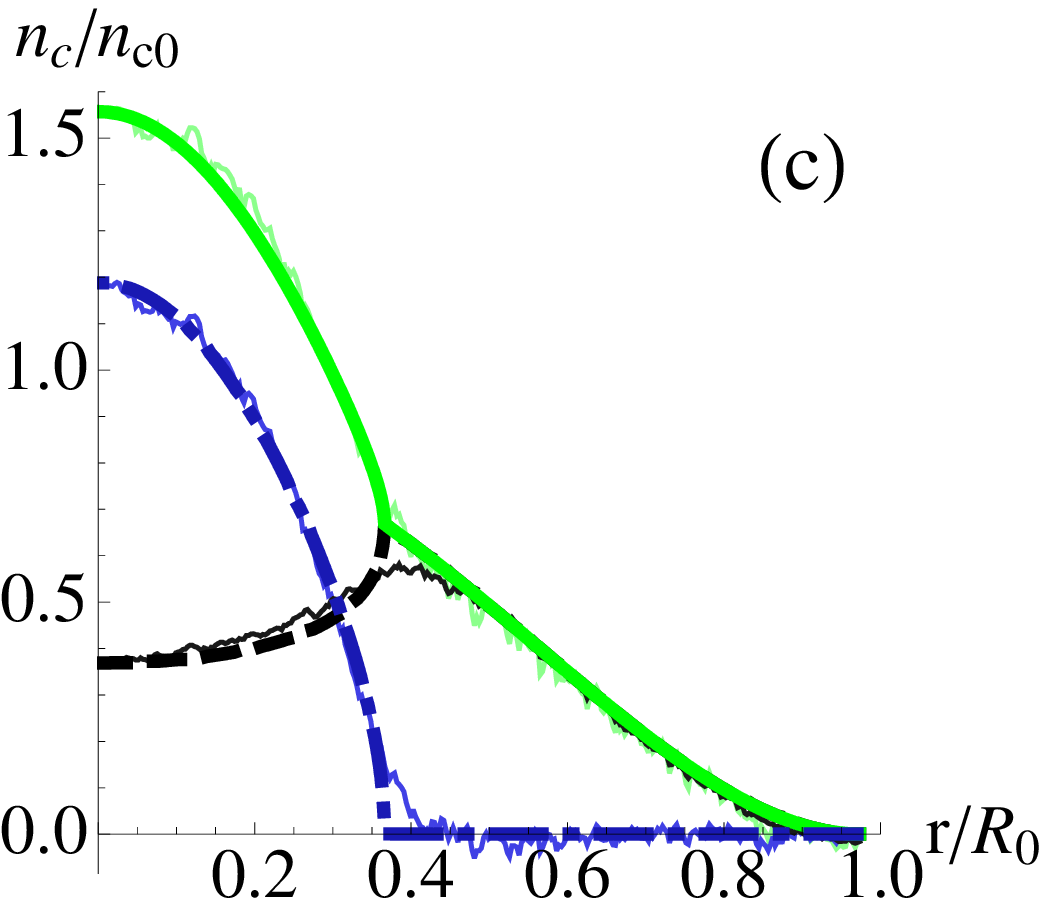} &
\includegraphics[angle=0, width=0.20\textwidth]{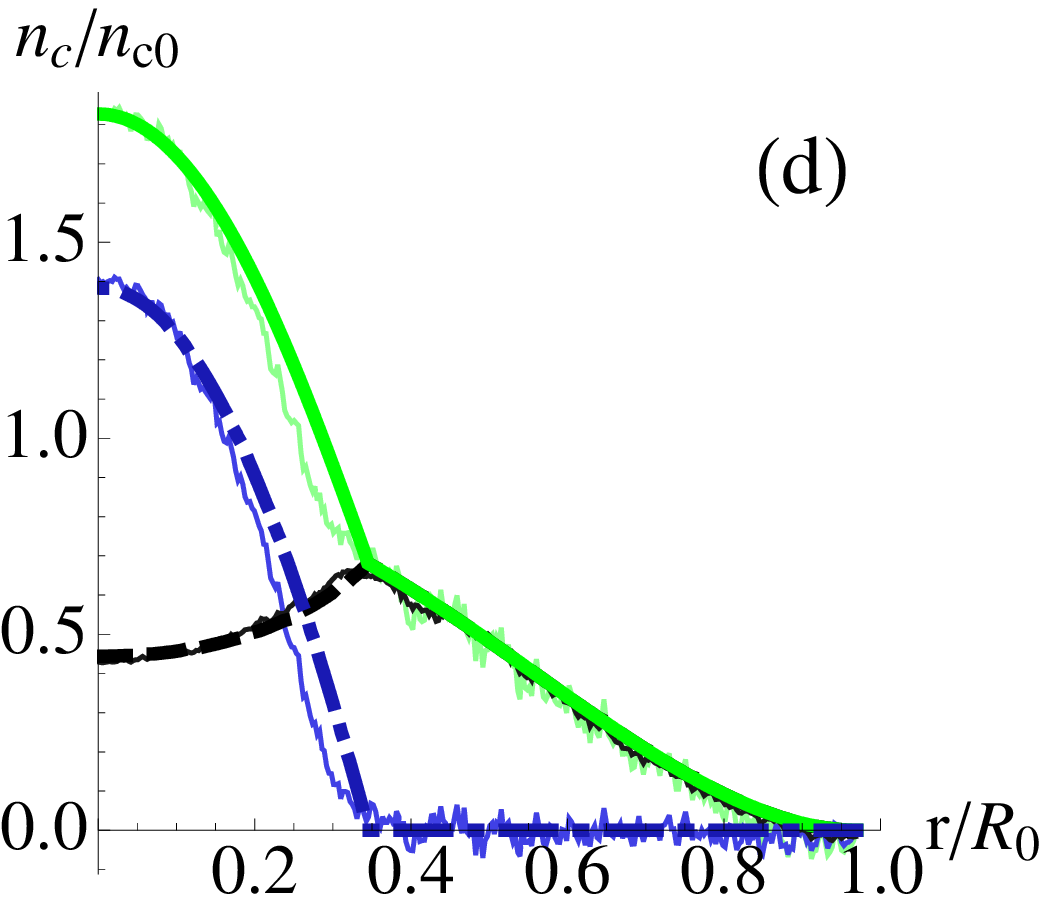} &
\includegraphics[angle=0, width=0.20\textwidth]{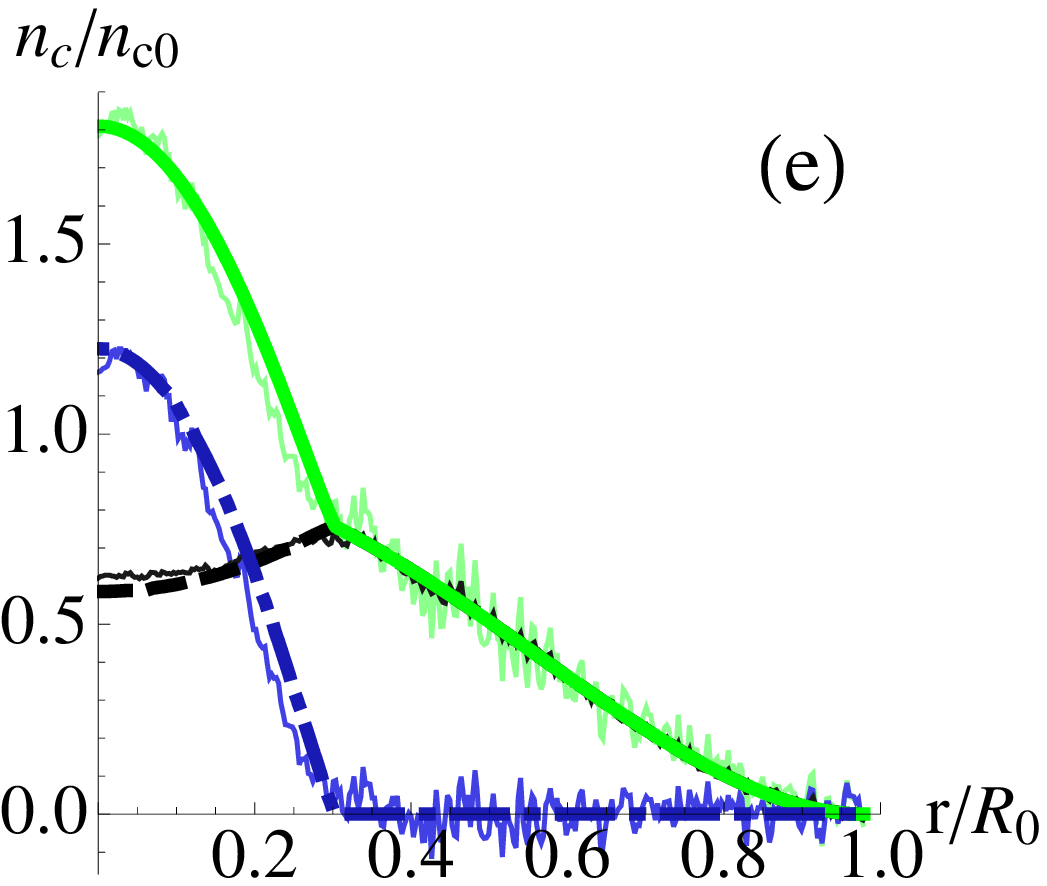} \\
\end{tabular}}
\caption{(color online). Comparison between calculated and experimental density (upper row) and column density (lower row) profiles for different values of polarization and interaction strength. Column (a) corresponds to $P=0.43$ at unitarity ($1/k_{F\uparrow}^0a=0$), column (b) to $P=0.52$ and $1/k_{F\uparrow}^0a=0.35$, column (c) to $P=0.57$ and $1/k_{F\uparrow}^0a=0.61$, column (d) to $P=0.61$ and $1/k_{F\uparrow}^0a=1.2$ and column (e) to $P=0.73$ and $1/k_{F\uparrow}^0a=1.6$. The theoretical profiles are plotted as in Fig.~\ref{fig2}. The corresponding experimental profiles are shown with thin lines. The reference density $n_0$, column density $n_{c0}$ and radius $R_0$ all refer to a non-interacting Fermi gas with $N_\uparrow$ particles. The experimental profiles are from Refs.~\cite{Shin08nat,Shin08}.}
\label{fig4}
\end{figure*}

\section{Results}
\label{Section3}

The shell structure of the atomic cloud for a given value of the polarization $P$ and interaction strength $1/k_{F\uparrow}^0a$ is schematically described by the phase diagram shown in Fig.~\ref{fig1}. The various regions denote different sequences of phases starting from the center of the trap. The most external shell is always composed by the majority spin-up atoms only (N$_{\text{FP}}$ phase). Outside the region delimited by the two (red) solid lines the sequence of phases corresponds to a continuous density profile, while inside it corresponds to a density profile which exhibits a jump marking the first order phase transition from the superfluid to the normal state. 

For large values of $P$ and close to the regime of resonant interaction, the system is normal and displays an internal shell of the N$_{\text{PP}}$ phase. The corresponding profile for both the spin-up and spin-down component is shown in column (a) of Fig.~\ref{fig2}. For the largest values of the interaction parameter $1/k_{F\uparrow}^0a$, corresponding to the leftmost region of Fig.~\ref{fig1}, the central part of the cloud is either a polarized SF$_{\text P}$ superfluid (see column (e) in Fig.~\ref{fig2}) or, for smaller polarizations, an unpolarized SF$_0$ superfluid surrounded by the SF$_{\text P}$ phase. In both cases the transition from the superfluid core to the N$_{\text{FP}}$ normal external shell is continuous and, within the picture of Bose-Fermi mixtures, corresponds to a vanishing density of composite bosons. The superfluid-normal transition is instead first order in the case of the other three shell sequences shown in Fig.~\ref{fig1}. These are: the SF$_0\to$N$_{\text{PP}}$ structure shown in column (b) of Fig.~\ref{fig2}, the SF$_0$+SF$_{\text P}\to$N$_{\text{FP}}$ shown in column (c) and the SF$_{\text P}\to$N$_{\text{FP}}$ shown in column (d). The jump occurring at the transition in the density profile of the minority spin component is shown in Fig.~\ref{fig2} with a vertical double arrow. 

The (blue) dashed line in Fig.~\ref{fig1} separates the region where an inner core of unpolarized SF$_0$ superfluid is present (below the line), from the region where the polarized SF$_{\text P}$ phase survives at the center of the cloud. The dotted line, instead, separates the region where the first order phase transition is from an unpolarized superfluid to a partially polarized normal gas (right of the line), from the region where it is from a polarized superfluid to a fully polarized normal gas (left of the line). The values of $P$ and $1/k_{F\uparrow}^0a$ for which a first order phase transition directly between the SF$_{\text P}$ and the N$_{\text{PP}}$ phase is predicted, would correspond to a very tiny region around the dotted line in Fig.~\ref{fig1}, smaller than in the homogeneous case (see Fig. 4 of Ref.~\cite{Pilati08}) and involving such small densities of the minority component to be experimentally irrelevant.

The structure of the cloud as a function of polarization and interaction strength can also be analyzed in terms of the radii of the various shells as shown in Fig.~\ref{fig3}. The radii of the spin-up and spin-down components, respectively $R_\uparrow$ and $R_\downarrow$, as well as the radii of the unpolarized superfluid core $R_c$ and of the superfluid component $R_{\text{SF}}$ are presented as a function of the polarization $P$ for different values of the interaction strength. Panel (a) of Fig.~\ref{fig3} shows the behavior of the radii at unitarity ($1/k_{F\uparrow}^0a=0$): the superfluid core exists here only if $P<0.77$, while for larger values of $P$ a partially polarized N$_{\text{PP}}$ gas is present for $R<R_\downarrow$. At $1/k_{F\uparrow}^0a=0.5$ [panel (b)] the inner part of the cloud is always superfluid and a N$_{\text{PP}}$ normal shell still survives surrounding the SF$_0$ superfluid core. For larger values of $1/k_{F\uparrow}^0a$ [panels (c) and (d)] the superfluid core is directly surrounded by the fully polarized N$_{\text{FP}}$ gas, but the radii $R_{\text{SF}}$ and $R_c$ are now different corresponding to the structure of an unpolarized SF$_0$ inner core surrounded by a polarized SF$_{\text P}$ external shell.

Finally in Fig.~\ref{fig4} we report the results of the comparison between the calculated density profiles and the ones measured in the recent experiment of Refs.~\cite{Shin08nat,Shin08}. We compare both density and column density profiles for different values of polarization and interaction strength. It is worth stressing that there are no free parameters in this comparison, once the values of $P$ and $1/k_{F\uparrow}^0a$ are extracted from the experimental data. The agreement is remarkable for all values of the interaction parameter from the unitary limit to the deep BEC regime, where it supports the picture of an interacting Bose-Fermi mixture~\cite{Shin08}. Small discrepancies, mainly visible as a broader density difference profile, are found close to the unitary limit [see panel (b) of Fig.~\ref{fig4}] and might be attributed to temperature effects~\cite{note}.

\section{Conclusions}
\label{Conclusions}

We have investigated the density profiles of trapped polarized Fermi mixtures at zero temperature. The starting point is the reliable determination of the phase diagram of uniform systems obtained in Ref.~\cite{Pilati08} using QMC methods, whose implications for trapped systems are derived by applying the local density approximation. The comparison with the profiles measured in experiments shows a good agreement from the unitary limit to the BEC regime, providing new physical insights into the strongly interacting polarized normal and superfluid phases recently realized with ultracold atoms.

\acknowledgments
We would like to thank Yong-il Shin for providing us with the experimental density profiles and for helpful discussions. We also thank S. Pilati for his help with the fitting procedures. This work, as part of the European Science Foundation EUROCORES Programme ``EuroQUAM-FerMix'', was supported by funds from the CNR and the EC Sixth Framework Programme. We also acknowledge support by the Ministero dell'Istruzione, dell'Universit\`a e della Ricerca (MIUR).


\begin{thebibliography}{20}
\bibitem{note} Notice that the experimental profiles shown in columns (b)-(e) of Fig.~\ref{fig4} are taken from Ref.~\cite{Shin08} and could correspond to a slightly higher temperature compared to the ones of column (a) which are taken from Ref.~\cite{Shin08nat}. See also the comparison in Fig. 4 of Ref.~\cite{Shin08nat}.
\end{thebibliography}
\end{document}